\documentclass[conference]{IEEEtran}

\IEEEoverridecommandlockouts

\usepackage{graphicx}
\usepackage{balance}  
\usepackage[show]{chato-notes}
\usepackage{paralist}
\usepackage{amsmath,amssymb}
\usepackage{mathtools}
\usepackage{spverbatim}
\usepackage{graphicx}
\usepackage{booktabs}     
\usepackage{multirow}
\usepackage[size=small,tableposition=top]{caption}     
\usepackage[scientific-notation=true]{siunitx}
\usepackage{bm}
\usepackage{float}
\usepackage{makecell}
\usepackage[export]{adjustbox}
\usepackage{url}

\newtheorem{mydefinition}{Definition}

\newtheorem{example}{Example}

\newtheorem{problem}{Problem}

\usepackage[vlined,linesnumbered,ruled,noend]{algorithm2e}     
\SetKw{Report}{report}
\SetKwFor{During}{during}{do}{}
\SetKwInOut{Output}{output}
\SetKwInOut{Input}{input}
\SetKw{Read}{read}
\SetKwComment{Comment}{//}{}
\SetKwComment{Section}{}{}
\SetCommentSty{textbf}

\SetKw{Or}{ or }
\SetKw{And}{ and }
\SetKw{Break}{ break }

\DeclareMathOperator*{\argmin}{arg\,min}
\DeclareMathOperator*{\argmax}{arg\,max}

\newcommand{\NPhard}{$\mathbf{NP}$-hard}

\newcommand{\define}{\ensuremath{{\,\leftarrow\,}}\xspace}
\DontPrintSemicolon
\newcommand{\spara}[1]{\smallskip\noindent{\bf #1}}

\newcommand{\squishlist}{
 \begin{list}{$\bullet$}
  {  \setlength{\itemsep}{0pt}
     \setlength{\parsep}{3pt}
     \setlength{\topsep}{3pt}
     \setlength{\partopsep}{0pt}
     \setlength{\leftmargin}{2em}
     \setlength{\labelwidth}{1.5em}
     \setlength{\labelsep}{0.5em}
} }
\newcommand{\squishlisttight}{
 \begin{list}{$\bullet$}
  { \setlength{\itemsep}{0pt}
    \setlength{\parsep}{0pt}
    \setlength{\topsep}{0pt}
    \setlength{\partopsep}{0pt}
    \setlength{\leftmargin}{2em}
    \setlength{\labelwidth}{1.5em}
    \setlength{\labelsep}{0.5em}
} }

\newcommand{\squishdesc}{
 \begin{list}{}
  {  \setlength{\itemsep}{0pt}
     \setlength{\parsep}{3pt}
     \setlength{\topsep}{3pt}
     \setlength{\partopsep}{0pt}
     \setlength{\leftmargin}{1em}
     \setlength{\labelwidth}{1.5em}
     \setlength{\labelsep}{0.5em}
} }
\newcounter{Lcount}
\newcommand{\squishenum}{
 \begin{list}{\arabic{Lcount}.}
  {  \usecounter{Lcount}
  	\setlength{\itemsep}{0pt}
     \setlength{\parsep}{3pt}
     \setlength{\topsep}{3pt}
     \setlength{\partopsep}{0pt}
     \setlength{\leftmargin}{2em}
     \setlength{\labelwidth}{1.5em}
     \setlength{\labelsep}{0.5em}
} }

\newcommand{\squishend}{
  \end{list}
}

\newtheorem{theorem}{Theorem}
\newtheorem{lemma}[theorem]{Lemma}

\def\BibTeX{{\rm B\kern-.05em{\sc i\kern-.025em b}\kern-.08em
    T\kern-.1667em\lower.7ex\hbox{E}\kern-.125emX}}
\begin{document}

\title{Adaptive Community Search in Dynamic Networks}
\author{
\IEEEauthorblockN{Ioanna Tsalouchidou}
\IEEEauthorblockA{\textit{Pompeu Fabra University} \\
Barcelona Spain \\
ioannatsalouchidou@gmail.com $\;\;\;$}
\and
\IEEEauthorblockN{Francesco Bonchi}
\IEEEauthorblockA{\textit{ISI Foundation}, Turin, Italy  \\
\textit{Eurecat}, Barcelona, Spain  \\
francesco.bonchi@isi.it}
\and
\IEEEauthorblockN{Ricardo Baeza-Yates}
\IEEEauthorblockA{\textit{Northeastern University} \\
Silicon Valley campus, CA, USA \\
rbaeza@acm.org}
}

\maketitle
\sloppy

\begin{abstract}
Community search is a well-studied problem which, given a static graph and a query set of vertices, requires to find a cohesive (or dense) subgraph containing the query vertices. In this paper we study the problem of community search in temporal dynamic networks.
We adapt to the temporal setting the notion of \emph{network inefficiency} which is based on the pairwise shortest-path distance among all the vertices in a solution. For this purpose we define the notion of \emph{shortest-fastest-path distance}: a linear combination of the temporal and spatial dimensions governed by a user-defined parameter.
We thus define the \textsc{Minimum Temporal-Inefficiency Subgraph} problem and show that it is \NPhard. We develop an algorithm which exploits a careful transformation of the temporal network to a static directed and weighted graph, and some recent approximation algorithm for finding the minimum Directed Steiner Tree. We finally generalize our framework to the streaming setting in which new snapshots of the temporal graph keep arriving continuously and our goal is to produce a community search solution for the temporal graph corresponding to a sliding time window.
\end{abstract}

\section{Introduction}
\label{sec:intro}
Community search, i.e., the problem of extracting cohesive subgraphs around a given set of vertices of interest, is a fundamental graph mining primitive which has received a great deal of attention~\cite{fang2019survey,HuangLX17}. If extracted from dynamic graphs, these substructures can help understanding the dynamics of the relationships that exist among these vertices. For instance, in a research collaboration network,  these structure can describe the dynamics of collaborations between a given group of researchers along time. As another example, the interactome, which is the set of molecular interactions in a cell, can be modeled as a network, in which the vertices are proteins and through their connections can perform biological functions. However, these connections are not constantly active, and therefore a dynamic analysis is more appropriate for understanding properly this complex network~\cite{Przytycka}.  Other application examples include events detection, friendship recommendation, control of infectious disease, semantic expansion, just to mention a few.
However,
\emph{surprisingly, little attention has been devoted to the community search problem in temporal networks.}
In this paper we propose a framework for adaptive community search in dynamic temporal networks
based on a model, defined for static graphs, by Ruchansky et al.~\cite{RuchanskyBGGK17}.
Given a static graph $G = (V,E)$ and a set of query vertices $Q \subseteq V$, the problem defined in~\cite{RuchanskyBGGK17} requires to find a set of vertices $S \supseteq Q$ such that its induced subgraph minimize \emph{network inefficiency}, a measure (formally defined in the next section) based on the pairwise shortest-path distance among all the vertices in the subgraph. The characterizing features of the proposal by Ruchansky et al. are that $(i)$ it is totally parameterless and $(ii)$ solutions can be disconnected: this way it allows to detect outliers in the query set, by letting such vertices disconnected from the others in the produced solution subgraph. The latter is an important feature, as community search is an explorative data analysis task,
in which we are given some query vertices which are \emph{suspected to be interesting}. However, real-world query-sets are noisy and likely to contain some vertices that are erroneously suspected of being related.
This feature is even more important in the temporal setting, in which vertices originally related might
become unrelated along time due to concept drift.

In order to generalize the notion of network inefficiency to temporal networks, the first step is to generalize the notion of shortest-path distance. As our goal is to use distances as a measure of cohesiveness, it is important to consider how the vertices connect through paths both in space (network structure) and in time (network evolution). For this purpose we adopt the notion of \emph{shortest-fastest-path distance} \cite{TsalouchidouBBL20}: a linear combination of the temporal and spatial dimensions governed by a user-defined parameter $\alpha \in [0,1]$. This definition generalizes both shortest and fastest path: by setting $\alpha=1$ we obtain shortest paths, while setting $\alpha =0$ we obtain fastest paths.
In general, depending on the application at hand, one can tune the parameter $\alpha$ to give more importance to the temporal dimension ($\alpha < 0.5$) or the spatial one ($\alpha >0.5$).

The next step is to define \emph{Temporal Network Inefficiency} for which we have to keep in consideration that, due to the intrinsic directionality of time, distances are no longer symmetric like in static undirected graphs. We finally can introduce the main problem studied in this paper which is the problem of extracting the minimum temporal-inefficiency subgraph problem  from a temporal graph.
We show that our problem is a genuine generalization of that Ruchansky et al.~\cite{RuchanskyBGGK17}: in fact,  for $\alpha = 1$ and in the case that the temporal graph is made of one single snapshot (it is thus not temporal), then the two problems exactly correspond.

Our proposed algorithm is structured in three phases. In the first phase the temporal graph is transformed in a static directed and weighted graph, by flattening all the temporal snapshots  in a unique static graph and by linking the various replicas of the same vertex in different timestamps, and by appropriately weighting the links. We show that the resulting transformed graph is such that the shortest-path distance between two vertices on this graph, corresponds to the shortest-fastest-path distance between the same vertices in the original temporal network. This property allows us to compute our solutions on this transformed graph. In the second phase we look for a \emph{connector} for the query vertices in $Q$, i.e., a set of vertices $S \supseteq Q$ such that in the induced subgraph all the vertices in $Q$ are connected, to the maximum possible extent.  For this task we extract minimum Directed Steiner Trees from transformed graph. In the third phase, we greedily reduce the solution $S$, returning as solution the subset which minimizes the temporal inefficiency.

Having an algorithm to extract the minimum temporal-inefficiency subgraph from a temporal graph defined over a fixed temporal window, our next step is to generalize our framework to the streaming setting. This is the setting in which new snapshots of the temporal graph keep arriving continuously and our goal is to produce, at each new timestamp, a community search solution for the temporal graph defined by a sliding temporal window of predefined size.
Since the network changes constantly in time, we expect that the communities evolve as well. Therefore, it is natural that the query set $Q$ is updated during the evolution, to keep in consideration the evidence of the key connections and most related vertices emerged during the previous time windows.

 We provide a streaming distributed implementation of our framework in Apache Spark and experiment on several real-world networks gathered by a proximity-sensing infrastructure recording face-to-face interactions in schools and co-authorship networks.

\section{Background and related work}
\label{sec:related}
%

Given a graph $G=(V,E)$ and a set of query vertices $Q \subseteq V$, the community search problem requires to  find a connected subgraph $H$ of $G$, that contains all query vertices $Q$ and that exhibits some nice properties of cohesiveness, compactness or density.
While optimizing for different objective functions, the bulk of this literature (see~\cite{fang2019survey,HuangLX17} for recent survey) shares a common aspect: the solution must be a \emph{connected} subgraph of the input graph containing the set of query vertices. Three recent approaches allow disconnected solutions in community search: allowing disconnected solutions is equivalent to allow some query vertices not to participate in the solution, thus being recognized as outliers. We call this version of the problem \emph{relaxed community search}.
Akoglu et al. \cite{akoglu2013mining} study the problem of finding \emph{pathways}, i.e., connection subgraphs for a large query set $Q$, in terms of the Minimum Description Length (MDL) principle.  According to MDL, a pathway is simple when only a few bits are needed to relay which edges should be followed to visit all of $Q$.
Given a graph $G$ and a query set $Q$, Gionis et al. \cite{gionisbump} study the problem of finding a connected subgraph of $G$ that has more vertices that belong to $Q$ than vertices that do not.  For a candidate solution $S$ that has $p$ vertices from $Q$ and $r$ not in $Q$, they define the \emph{discrepancy} of $S$ as a linear combination of $p$ and $r$, and study the problem of maximizing discrepancy. They show that the problem is \NPhard\ and develop efficient heuristic algorithms.

The work which we extend to the case of temporal networks, is that of Ruchansky et al. \cite{RuchanskyBGGK17} which studies, in the static graph setting, the problem of extracting the
\emph{minimum inefficiency subgraph}. Let $d_{G[S]}(u,v)$ denote the shortest-path distance
between a pair of vertices $u,v \in S$, computed in the subgraph induced by $S$, then the problem is defined as follows.

\smallskip 

\begin{problem}[\textsc{Min-Inefficiency-Subgraph} \cite{RuchanskyBGGK17}]\label{prob:mis}
\emph{Given an undirected graph $G = (V,E)$ and a query set $Q \subseteq V$, find the not necessarily connected subgraph $H^*$ minimizing the network inefficiency:
$$H^* = \argmin_{G[S] : Q \subseteq S \subseteq V}\sum_{u,v \in S, u\neq v}{1-\frac{1}{d_{G[S]}(u,v)}}.$$}
\end{problem}

The intuition at the basis of this definition is that all-pairs shortest-path distances provide a good measure of how cohesive is a subgraph \cite{ruchansky2015minimum,RuchanskyBGGK17}.
However, one issue with shortest-path distance is that it is enough to have one disconnected vertex to have an infinite measure.
A simple yet elegant workaround to this issue, is to use the reciprocal of the shortest-path distance, under the convention that $\infty^{-1} = 0$. Therefore, a disconnected pair of nodes produces $1 - \frac{1}{\infty} = 1$ inefficiency. On the opposite, a pair of adjacent vertices produces $1 - 1 = 0$ inefficiency. A pair of vertices at distance 2 produces  $1/2$ inefficiency, a pair at distance 3 produces $2/3$ inefficiency, and so on.

Ruchansky et al.  show that Problem \ref{prob:mis} is  \NPhard\ and develop an efficient greedy algorithm. They also show that the minimum inefficiency subgraph exhibits some nice properties such as the fact of producing small cohesive solutions, the fact of being able to detect outlier query vertices (by leaving them disconnected in the produced solution), and the fact of being able to recognize when the query vertices belong to different communities.
Ruchansky et al. \cite{RuchanskyBGGK17} also provide an empirical (qualitative and quantitative)
comparison with~\cite{akoglu2013mining,gionisbump}.
%

\spara{Temporal paths and Steiner trees on dynamic networks.}  
Given a dynamic network, where edges are timestamped, temporal paths are paths in the graph structure, along the temporal dimension. In particular, temporal paths must be
\emph{time-respecting}, that is, edges along a path appear in non-decreasing time~\cite{KempeKK02}. Bui-Xuan et al.~\cite{XuanFJ03} study interesting paths as either shortest, fastest or foremost journeys. Wu et al.~\cite{Wu:2014} propose more efficient methods to compute these paths in both streaming and transformed graph models. Recently, some parallel and distributed algorithms for computing temporal paths~\cite{wu2016efficient, ni2017parallel} have been proposed.

A related problem that we use in our algorithm  is the computation of the minimum Directed Steiner Tree in static graphs for which a first approximation algorithm was proposed by Charikar et al.~\cite{CharikarCCDGGL99}. Recently Huang et al.~\cite{HuangFL15} have studied the problem of minimum spanning trees in temporal graphs and propose two definitions based on the optimization of time and cost. They also propose an improved approximation algorithm for the minimum Directed Steiner Tree problem for temporal graphs, using a graph transformation from temporal to static graphs, similar to the one we use in Section 4.

\section{Temporal network inefficiency}
\label{sec:problem}
We are given a continuous stream of timestamped edges $(u,v,t)$, where $u,v \in V$ are vertices and $t$ is a timestamp from a potentially infinite domain $T$. We can represent a dynamic graph as a sequence of graphs, one for each timestamp, defined over the same set of vertices  i.e., $\mathcal{G} =\langle G_0, G_1, ..., G_t,...\rangle$ with $G_t = (V,E_t)$ and $E_t=\{(u,v,t)\}$. Given a temporal interval $[i,j]$ we denote $G_{[i,j]}$ the projection of $\mathcal{G}$ over $[i,j]$.
A temporal path between a pair of vertices $v,u \in V$ is a time-respecting sequence of edges, i.e., $$p(u,v) = \{(u=v_0,v_1,t_0),(v_1,v_2,t_1),\ldots,(v_n,v_{n+1}=v,t_n)\}$$  such that $\forall i \in [1,n]$ it holds that $t_{i-1} \leq t_{i}$.

\smallskip

\begin{figure}[t!]
 \begin{minipage}{0.18\textwidth}
   \includegraphics[width =\linewidth ]{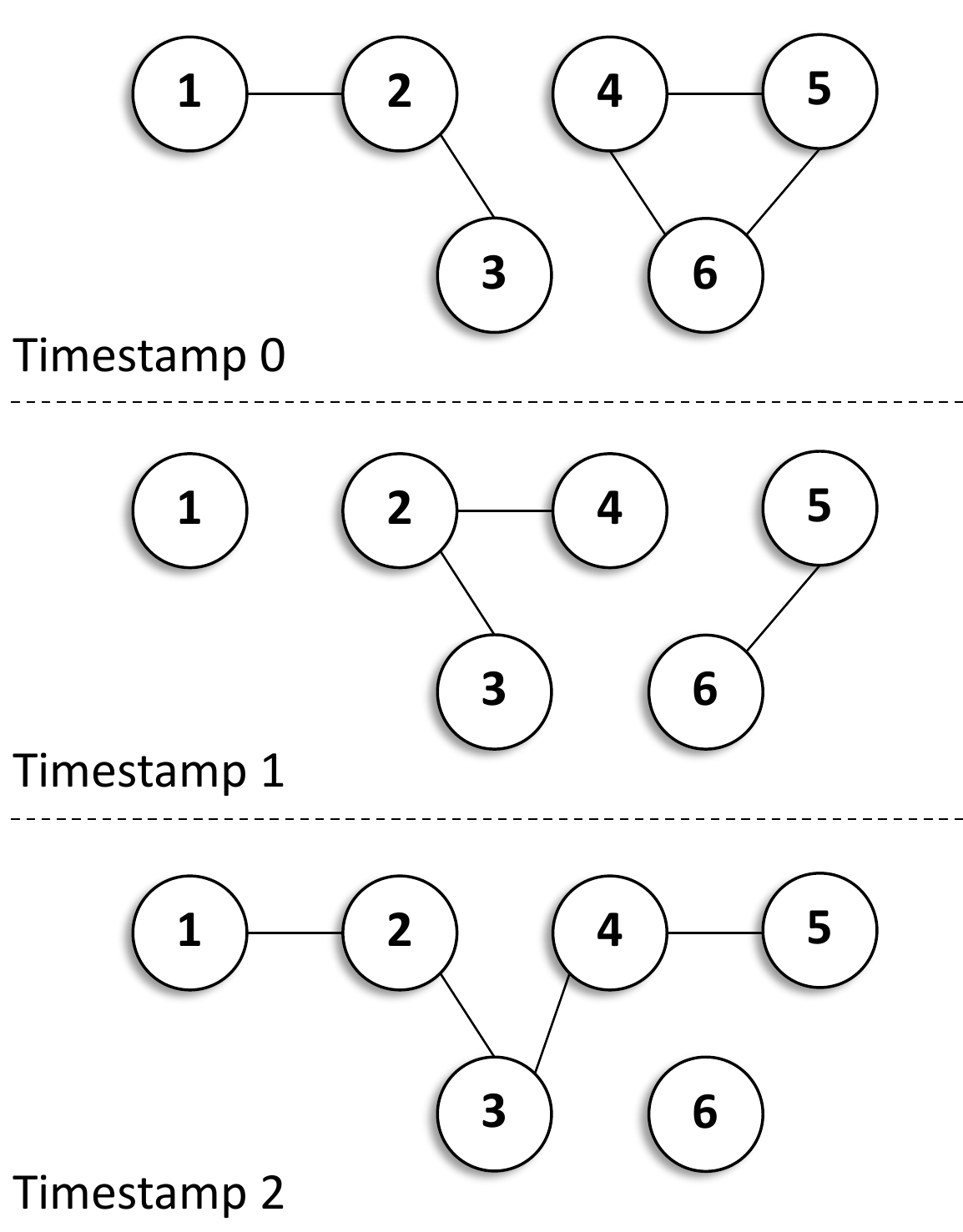}
 \end{minipage}
  \begin{minipage}{0.26\textwidth}
 \begin{scriptsize}
  \begin{tabular}{r l r r}
    \toprule
$u$&$v$	&  $d_{G_{[0,2]}}(u,v)$	& $d_{G_{[0,2]}}(v,u)$ \\
    \midrule
1&2				&	$\alpha$		&	$\alpha$				\\
1&	3				&	$2 \alpha$	&	$2 \alpha$			\\
1&4		&	$min(3\alpha,\alpha+1)$	&	$min(3\alpha,\alpha+1)$			\\
1&5			&	$min(4\alpha,\alpha+2)$	& $min(4\alpha,\alpha+2)$			\\
1&6			&	$\infty$		&	$\alpha +2$		\\
2&3					&	$\alpha$		&	$\alpha$		\\
2&4		&	$\alpha$	&	$\alpha$	\\
2&5			&	$min(3\alpha,\alpha+1)$	&	$min(3\alpha,\alpha+1)$	\\
2&6			&	$\infty$		&	$\alpha+1$	\\
3&4			&	$\alpha$		&	$\alpha$		\\
3&5			&	$2 \alpha$	&	$2 \alpha$	\\
3&6			&	$\infty$		&		$min(2\alpha +1, 2)$				\\
4&5				&	$\alpha$		&	$\alpha$		\\
4&6			&	$\alpha$		&	$\alpha$		\\
5&6			&	$\alpha$		&	$\alpha$		\\
    \bottomrule
  \end{tabular}
   \end{scriptsize}
 \end{minipage}

	\caption{An example dynamic graph on three timestamps, and the shortest-fastest-path distance between any pair of vertices.}\label{fig:toy_example}
\end{figure}

\begin{example}
\emph{
A time-respecting path between vertex 6 and vertex 1 in Figure~\ref{fig:toy_example} is  $p(6,1)= \{(6,4,0),(4,2,1),(2,1,2)\}$. However, there exists no time-respecting path from 1 to 6. It is worth noticing that in a dynamic graph, even if undirected, due to the notion of temporal path, distance is not symmetric.}
\end{example}

\smallskip

When dealing with temporal dynamic graphs, one can use different characteristics to define the interestingness of a path between two vertices. In fact, besides the usual spatial definition of shortest-path distance based on the number of intermediate vertices, one can also consider the temporal duration of the path itself. For instance, Wu et al.~\cite{Wu:2014} study four different types of interesting paths over temporal graphs within a time window: \begin{inparaenum}[(1)]\item earliest-arrival path, \item latest-departure path, \item fastest path, and \item shortest path\end{inparaenum}.
In this paper we adopt a linear combination of the temporal and spatial distance\footnote{Throughout this paper by ``spatial'' distance we mean the usual shortest-path distance defined as the minimum number of hops between two vertices.}, governed by a parameter $\alpha  \in [0,1]$.


\smallskip

\begin{mydefinition}[Shortest Fastest Path \cite{TsalouchidouBBL20}]\label{SFP}
\emph{Given a user-defined parameter $\alpha \in [0,1]$ we define as shortest fastest path (SFP) between a pair of vertices $u,v \in V$ in a dynamic graph, a valid temporal path $p(u,v) = \{(u,v_1,t_0),\ldots ,(v_n,v,t_n)\}$ minimizing the cost:}
\begin{equation}\label{function:objective}
 \mathcal{L}(p) = \alpha|p(u,v)| + (1-\alpha)(t_n - t_0)
\end{equation}
\end{mydefinition}

This definition generalizes both shortest and fastest path notions. In fact by setting $\alpha=1$ we obtain shortest paths, while setting $\alpha =0$ we obtain fastest paths.
In general, depending on the application at hand, one can tune the parameter $\alpha$ to give more importance to the temporal dimension ($\alpha < 0.5$) or the spatial one ($\alpha >0.5$). The parameter $\alpha$ can also be tuned in such a way to favor one dimension, but using the other dimension for tie-breaking among equivalent paths in the first dimension. For instance, by setting $\alpha$ to a small positive quantity $\epsilon$, the temporal paths that we obtain by Eq.~(\ref{function:objective}) correspond to \emph{fastest paths} with the minimum number of intermediate hops. Similarly, if we set $\alpha= 1-\epsilon$, Eq.~(\ref{function:objective}) will return the shortest paths, and among all shortest paths, the one employing  the minimum number of timestamps.

\smallskip

\begin{mydefinition}[Shortest-Fastest-Path Distance]
\emph{The shortest-fastest-path distance between two vertices $u,v$ is  the cost of the shortest fastest path between the two vertices, i.e.,
$d_G(u,v) = \mathcal{L}(p^*(u,v))$, where $p^*(u,v)= \argmin \mathcal{L}(p(u,v)).$
}
\end{mydefinition}

\smallskip

\begin{example}
\emph{Consider again the example in Figure~\ref{fig:toy_example} and suppose we want to go from vertex 1 to 4.
We have a short path through vertex 2 over two timestamps, i.e., $\{(1,2,0),(2,4,1)\}$ or alternatively we have a bit longer path which is faster, as it spans only one timestamp: i.e., $\{(1,2,2),(2,3,2),(3,4,2)\}$.
The cost of the former path is $\alpha +1$, while the cost of the latter is $3\alpha$. An $\alpha > 0.5$ would favor the shortest path, while an $\alpha < 0.5$ will favor the fastest path. \label{ex2}}
\end{example}

\smallskip

Two simple observations are in place. First, in a dynamic graph, even if undirected, due to the notion of time-respecting path, the distance is no longer symmetric (nor it is reachability). Second, while distance is static graphs is usually in $[1,\infty]$, with our definition of shortest-fastest-path distance the minimum value is $\alpha$. With these two observations in mind, we are now ready to extend the notion of network inefficiency by Ruchansky et al. \cite{RuchanskyBGGK17}, to the case of temporal dynamic graphs.

\smallskip

\begin{mydefinition}[Temporal Network Inefficiency]\label{definition:TemporalInefficiency}
\emph{Given a temporal graph $G_{[0,t]}$ defined over a set of vertices $V$ and $t+1$ timestamps, and given
a user-defined parameter $\alpha \in [0,1]$, the network inefficiency of $G_{[0,t]}$ is defined as
 \[ \mathcal{I}(G_{[0,t]}) = \sum_{u,v \in V, u\neq v}{\frac{(1-\frac{\alpha}{d_{G_{[0,t]}}(v,u)})+(1-\frac{\alpha}{d_{G_{[0,t]}}(u,v)})}{2}}.\]	}
\end{mydefinition}

Definition~\ref{definition:TemporalInefficiency} differs from the notion of static network inefficiency in  two main points. The first is that it uses reciprocal of the distance between two vertices, multiplied by $\alpha$.
This is to keep the values of the inefficiency in $[0,1]$, as in the static case, since as we mentioned before, the minimum length of a temporal path is $\alpha$.
The second change, is due to the asymmetry in the distance of the temporal path between two vertices. Therefore, for each pair of vertices, we calculate the inefficiency introduced in the network by considering the distance in both directions of the path.

\begin{example}
\emph{Note that two vertices that are directly connected in at least a timestamp (as for instance vertices 1 and 2 in Figure~\ref{fig:toy_example}) have a distance of $\alpha$, and thus they do not add any temporal inefficiency to the network. Let us now consider the pair of vertices (1,6).
We already saw that there is no time-respecting path from vertex 1 to 6: thus this direction provides maximum inefficiency of 1. In the other direction we have the temporal path $p(6,1)= \{(6,4,0),(4,2,1),(2,1,2)\}$ of lenght 3 and spanning 3 timestamps (thus, with a distance of $\alpha +2$).
Assuming $\alpha = 0.5$, temporal inefficiency brought by the pair of vertices (1,6) is
$$
\frac{(1-\frac{0.5}{\infty})+(1-\frac{0.5}{2.5})}{2} = \frac{1 + 0.8}{2} = 0.9
$$
which is close to the maximum inefficiency of $1$.}
\end{example}

\smallskip

We are now ready to formalize the problem studied in this paper. We do this in steps: first we define the problem of finding the minimum temporal-inefficiency subgraph in a single time window $W$; then we extend the framework to deal with a sliding window, on a potentially infinite stream of graphs, in an adaptive way.

Let us consider the temporal graph $G_W = (V, E_W)$ defined over the time interval $W = [t_i-(|W|-1),t_i]$ with $t_i \in T$. Given a set of vertices $S \subseteq V$, we denote $G_{W}[S]$ the temporal subgraph induced by $S$, i.e.,  $G_W[S] = (S, \{(u,v,t) \in E_W | u \in S \wedge v \in S  \})$.
Given a set of query vertices $Q$, the minimum temporal-inefficiency subgraph problem requires to find a set of vertices $S$ containing $Q$, such that its induced subgraph minimizes the temporal inefficiency (Definition \ref{definition:TemporalInefficiency}).

\smallskip

\begin{problem}[\textsc{Minimum Temporal-Inefficiency Subgraph}\label{problem1}]
\emph{Given a temporal graph $G_W = (V, E_W)$, a parameter $\alpha \in [0,1]$ and a set of query vertices $Q \subseteq V$, find the minimum temporal-inefficiency subgraph of $G_W$:
\[H^* = \argmin_{G_W[S]:Q\subseteq S \subseteq V} \mathcal{I}(G_W[S]).\]}
\end{problem}

\smallskip

We observe that when the temporal graph $G_W$ is made of only one snapshot (i.e., $|W|=1$) and $\alpha = 1$,
Problem \ref{problem1} exactly coincides to the \textsc{Min-Inefficiency-Subgraph}
on standard graphs (Problem \ref{prob:mis})
of  Ruchansky et al.
\cite{RuchanskyBGGK17}.
As Problem \ref{prob:mis} is \NPhard\ and it is a special case of Problem \ref{problem1}, it holds that also the \textsc{Minimum Temporal-Inefficiency Subgraph} problem is \NPhard.

Problem \ref{problem1} above produces one solution subgraph for an input temporal graph defined over a fixed temporal interval.
We next consider the \emph{sliding window} setting, where the temporal graph of interest is defined
by a temporal interval $W=[t-(|W|-1),t]$ with $t\in T$ of predefined length $|W|$, which keeps sliding continuously along time. At every new timestamp the window $W$ is updated by adding a new snapshot of the graph and removing the most obsolete one. Therefore, at every new timestamp $t\in T$, we have a new window $W$ which, in turns, defines a new temporal graph $G_W$, and we want to solve a new instance of the Problem~\ref{problem1} outputting a new community. This enables the analysis of how the community around a query set of vertices $Q$ changes along time.

\spara{Adaptive Query Set. }
One of distinguishing feature that our framework inherits from \cite{RuchanskyBGGK17} is the fact of producing solutions which are not necessarily connected: if a vertex in the query set $Q$ results to be
 not so related to the others, it can be treated as an \emph{outlier} by leaving it disconnected  in the solution. This flexibility is an even more interesting feature in the context of temporal graphs analysis as it allows to deal with \emph{concept drift}: something which is interesting and relevant at the beginning of the temporal domain under analysis, can become less and less relevant as time progresses.

Based on this observation and exploiting the flexibility of our framework, we propose the \emph{adaptive query set} $Q$  mechanism, as a way to deal with concept drift. This is achieved by means of two simple update rules that have the faculty of changing $Q$ amid evidence emerging from the analysis.
The two update rules are as follows:
\squishlist

\item if a vertex $v \in Q$ remains disconnected in the solution $H_t$ for a number of consecutive timestamps $t$ larger than a user-defined threshold $\lambda_{out} \in \mathbb{N}^+$, then $v$ is removed from $Q$;

\item if a vertex $v \in V \setminus Q$ appears in the solution $H_t$ for a number of consecutive timestamps $t$ larger than a user-defined threshold $\lambda_{in} \in \mathbb{N}^+$, then $v$ is added to $Q$.
\squishend

The rationale for these update rules is that in an explorative data analysis task,
real-world query-sets are likely to contain some vertices that are erroneously suspected of being related, or that become unrelated along time due to concept drift. Similarly, a vertex which is added constantly to the solution, as it helps reducing the distance among the vertices in $Q$, can be considered related to the others in $Q$ and added to the query set for future monitoring.

\section{Algorithms}
\label{sec:algorithms}
We next present our method for solving Problem \ref{problem1}, i.e., extracting the minimum temporal-inefficiency subgraph from a temporal graph $G_W = (V, E_W)$. Later we discuss the sliding window setting and its streaming-distributed implementation.

Our method is structured in three phases:

  \spara{Phase 1: Graph Transformation --} the temporal graph $G_W$ is transformed in a static directed and weighted graph $G'$, by flattening all the snapshots $G_t$ with $t \in W$ in a unique graph linking the various replicas of the same vertex in different timestamps, appropriately weighting these edges and the original edges, and adding some source and sinks dummy vertices. We show that the resulting transformed graph $G'$ is such that the cost $\mathcal{L}(p)$ of the shortest-fastest path between two vertices in $G_W $ can be computed as shortest-path distance between the corresponding source and sink vertices in $G'$.

   \spara{Phase 2: Temporal connector --} during the second phase we compute a \emph{connector} for the query vertices in $Q$, i.e., a set of vertices $S \supseteq Q$ such that in the induced subgraph $G_W[S]$ all the vertices in $Q$ are connected, to the maximum possible extent.  For this task we extract many Directed Steiner Trees from transformed graph $G'$.

\spara{Phase 3:  Minimum Temporal-Inefficiency Subgraph --} in the third phase, following \cite{RuchanskyBGGK17}, we greedily reduce the solution $S$, returning as solution the subset which minimizes the temporal inefficiency.

\smallskip

We next present each phase in details, then we discuss sliding window setting and the adaptivity of the query set.

\subsection{Phase 1: Graph Transformation}
Given a temporal graph $G_W = (V, E_W)$ and the parameter $\alpha \in [0,1]$ as in Problem \ref{problem1}, we build a static directed weighted graph  $G'(V',E',\ell)$, where $\ell: E' \rightarrow [0,1]$ is the edge weighting function, as follows:
\squishlist
\item \textbf{Vertices:} for each $t \in W$ and each $v \in V$ we create a new vertex having as id the pair vertex-timestamp $(v,t)$, that is $V' = \{(v,t): t \in W, v \in V\}$.

\item \textbf{Edges:} for each $v \in V$  and each pair of consecutive timestamps $t_i, t_{i+1} \in W$ with we create a directed edge $((v,t_i),(v,t_{i+1}))$ with weight $1 - \alpha$. The edges in $E_W$, are instead assigned a weight of $\alpha$.
\squishend

\begin{figure}[t!]
  \centering
  \includegraphics[width=.5\linewidth]{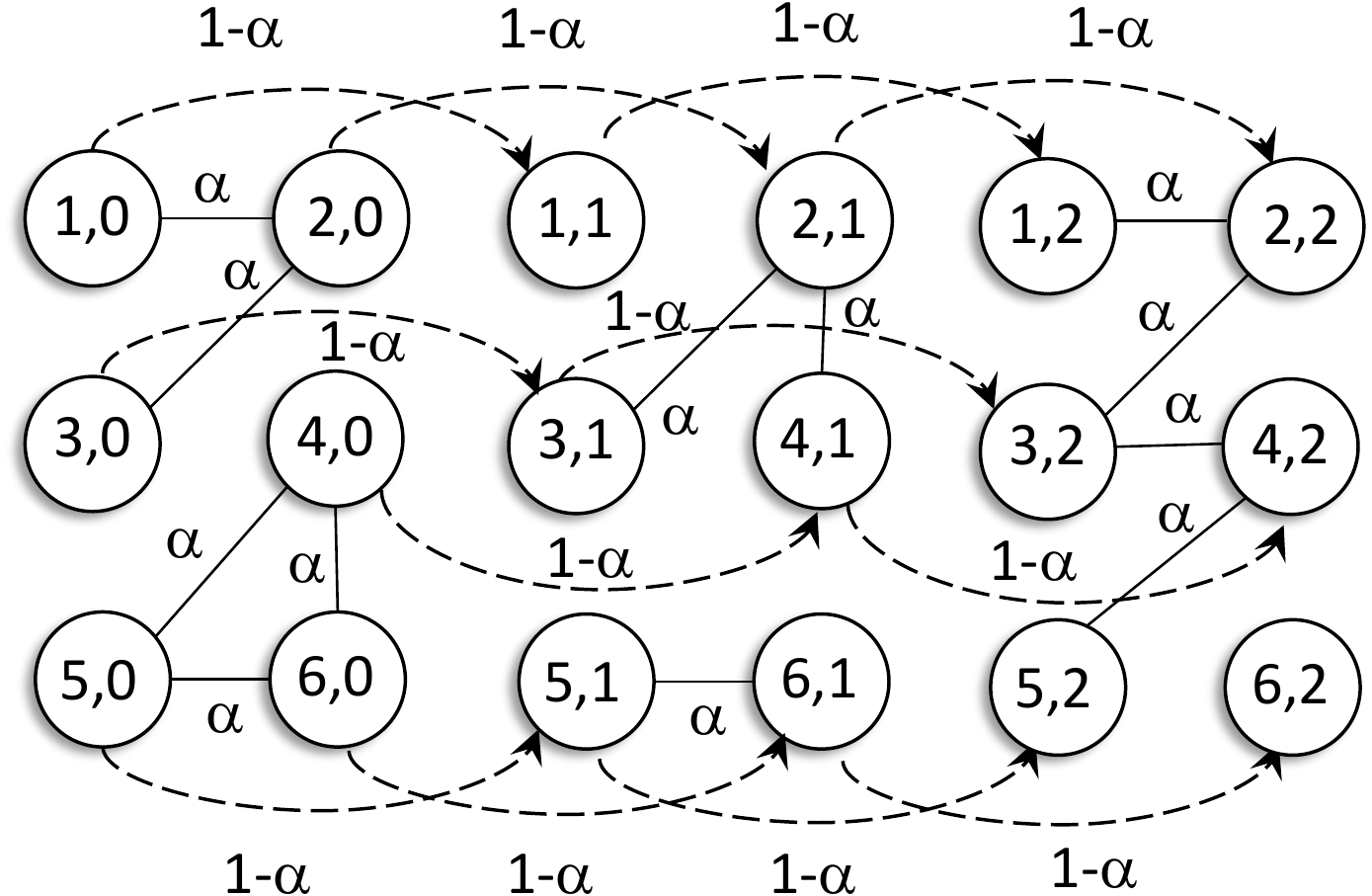}
  \caption{The transformed graph for the example temporal graph $G_{[0,2]}$ from Figure \ref{fig:toy_example}.}\label{fig:example3b}
\end{figure}

Figure \ref{fig:example3b} shows the transformed graph for the example temporal graph $G_{[0,2]}$ from Figure \ref{fig:toy_example}.

 Let us consider now a pair of vertices $(u,v) \in V \times V$, and let us define $P(u,v) = \{p((u,t_i),(v,t_j)) | t_i, t_j \in W \}$ the set of all paths from any replica of $u$ to any replica of $v$ in the transformed graph $G'$. Let us also denote $\ell(p)$ the length of one such path $p$, i.e., the sum of the weights of the edges in the path. Thanks to the edge labeling in $G'$, the following (straightforward) lemma holds.
\begin{lemma}
\emph{A path $p$ on the transformed graph $G'$  from a replica of $u$ to a replica of $v$, corresponds to a valid temporal path $p'$ from $u$ to $v$ in the temporal graph  $G_W = (V, E_W)$, and the length of $p$ in $G'$ corresponds to the cost of $p'$ in  $G_W$, i.e.,  $\ell(p) = \mathcal{L}(p').$\label{lemma1}}
\end{lemma}

\smallskip

\begin{example}
\emph{On the temporal graph $G_{[0,2]}$ from Figure \ref{fig:toy_example}, consider the time-respecting path $p(6,1)= \{(6,4,0),(4,2,1),(2,1,2)\}$ . Figure~\ref{fig:example3b} reports the same path on the transformed graph $G'$.
 We can observe that the cost of the path is $\alpha + 2$ as dictated by Eq.(\ref{function:objective}) in Definition \ref{SFP}.}
\end{example}

From Lemma \ref{lemma1} it follows that to compute the shortest fastest path (SFP) between a pair of vertices $u,v \in V$ we can simply find the shortest path between any replica of $u$ to any replica of $v$ on the transformed graph $G'$. Moreover it holds that the length of such shortest path on $G'$ corresponds to the shortest-fastest-path distance $d_{G_{[0,2]}}(u,v)$. In order to consider all possible shortest paths on $G'$ from any replica of a vertex $u$ to any replica of a vertex $v$ we create two dummy vertices: a dummy source with id $u$ and a dummy sink with id $(v,-1)$. By starting Dijkstra's algorithm from dummy source vertex $u$ we can find the shortest path to sink $(v,-1)$, which corresponds to the SPF from $u$ to $v$ on $G_{[0,2]}$.

\begin{figure}[t!]
  \centering
  \includegraphics[width=.45\linewidth]{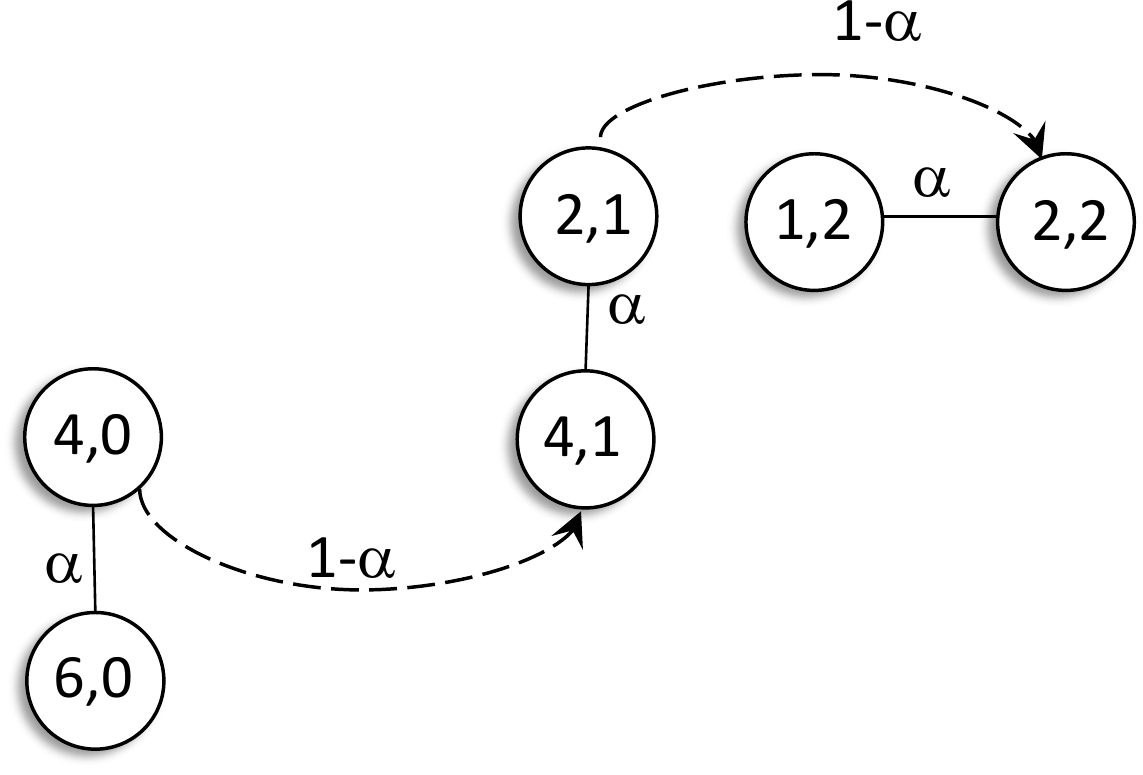}
  \caption{The temporal path $p(6,1)= \{(6,4,0),(4,2,1),(2,1,2)\}$ from Example 1 reported on the transformed graph $G'$. We can see that the path has a cost of $\alpha + 2$.}\label{fig:example3c}
\end{figure}

\smallskip

\begin{example}
\emph{Consider again the running example on $G_{[0,2]}$ (Figure~\ref{fig:toy_example}), and suppose we want to compute the shortest-fastest-path distance $d_{G_{[0,2]}}(1,4)$. On $G'$ augmented with dummy source vertex $1$ and sink $(4-1)$, we compute all shortest paths from the source to the sink.}

\begin{figure}[h!]
\vspace{-3mm}
  \centering
  \includegraphics[width=.5\linewidth]{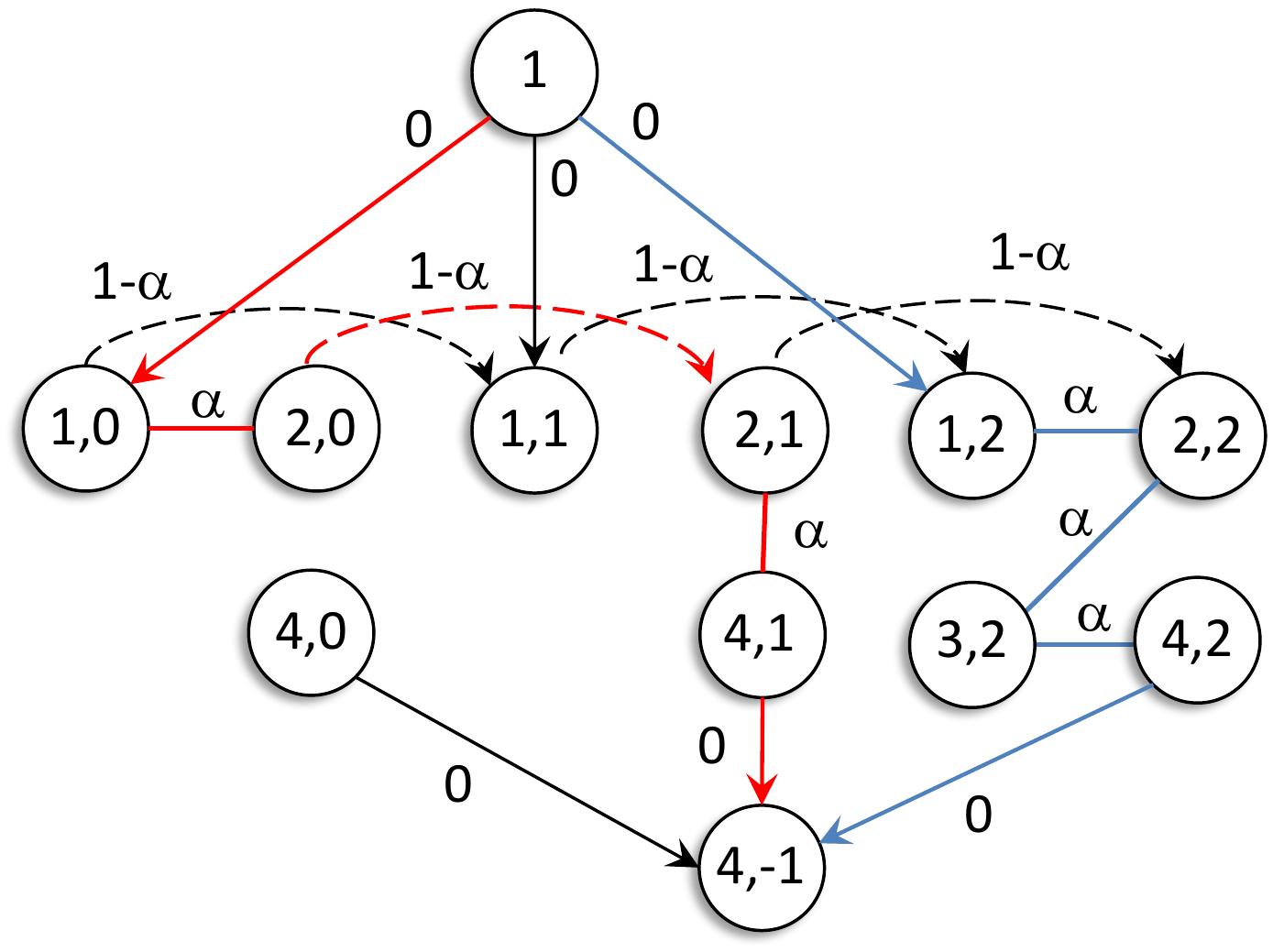}
  \caption{Possible shortest-fastest paths from 1 to 4: the red one (if $\alpha > 0.5$) or the blue one (if $\alpha < 0.5$).}\label{fig:example3d}
\end{figure}

\emph{In Figure \ref{fig:example3d} we report the two paths, previously discussed in Example \ref{ex2}, $\{(1,2,0),(2,4,1)\}$ (in red) and $\{(1,2,2),(2,3,2),(3,4,2)\}$ (in blue). We can see how their length computed over $G'$ are $\alpha +1$ and $3\alpha$, respectively. Which of the two is the shortest-fastest path depends on the value of $\alpha$, as discussed in Example \ref{ex2}. We observe that it holds
that $d_{G_{[0,2]}}(1,4) = min(3\alpha,\alpha+1)$ as reported in Figure \ref{fig:toy_example}.}
\end{example}

 \smallskip

The phase 1 of our method, graph transformation, is described in lines \ref{line:begin_transformation}-\ref{line:end_transformation} of Algorithm~\ref{algorithm:MIS}.

\subsection{Phase 2: Temporal Connector} In the second step we aim at building a \emph{connector} for the set of query vertices $Q$, i.e., a set of vertices $S \supseteq Q$ such that in the induced subgraph $G_W[S]$ all the vertices in $Q$ are connected, to the maximum possible extent. In fact, due to the requirement of time-respecting paths, not all vertices in $Q$ are necessarily reachable from all the other vertices in $Q$.

We deal with this task by building \emph{Directed Steiner Trees}
 (DST)~\cite{CharikarCCDGGL99} on the transformed graph $G'$ augmented with the dummy source and sink vertices for all $v\in Q$.

 \smallskip

\begin{mydefinition}[Minimum Directed Steiner Tree]
\emph{Given a directed weighted static graph $G=(V,E,\ell)$ where $\ell$ is the edge-weighting function.
Given a set of terminal vertices $X \subset V$ and a root vertex $r \in V$, the minimum directed Steiner tree problem (DST) ask to find the minimum cost tree which connects the root $r$ to each terminal $v \in X$. In the case in which not all the terminals are reachable from the root, then the minimum cost achievable is $+\infty$. In this case the problem requires to return a tree that covers as much as possible of $X$, and among the trees that do so, the one of minimum cost.}
\end{mydefinition}

 \smallskip

More precisely, for each $r \in Q$ we search for the minimum DST using the dummy source vertex $r$ as the root and the dummy sink vertices $\{(u,-1) | u \in Q \setminus \{r\} \}$ as terminals.
For computing the minimum DST on the transformed graph $G'$ augmented with the dummy source and sink vertices, we exploit the recent approximation algorithm by Huang et al.~\cite{HuangFL15} appropriately modified to take care of the disconnected cases.
It is important to note that the method in~\cite{HuangFL15} requires in input the transitive closure of the graph $G'$ which might be a computational bottleneck. In Algorithm~\ref{algorithm:MIS},
lines \ref{line:transitive_closure} we report the transitive closure computation as it was a pre-processing step. However, later in Section \ref{sec:distributed} we discuss a way we use to reduce the computational overhead.

Let us denote $MDST(G',r,Q \setminus \{r\})$ such minimum DST, or simply $MDST(r)$. We also overload the weighting function $\ell$ to denote the total cost of the tree. i.e., $\ell(MDST(r))$.
Our method selects $$r^* = \argmin_{r \in Q} \ell(MDST(r)).$$
If $\ell(MDST(r)) < \infty$ then we are done: we select as connector in the original temporal graph $G_W$ the set of vertices whose replicas are involved in $MDST(r)$ computed over the transformed graph $G'$, i.e., $S = \{ u \in V| (u,t_i) \in MDST(r) \}$. Otherwise, if $\ell(MDST(r)) = \infty$, then there are disconnected terminals. Let us denote the set of disconnected terminals $\overline{Q}$. Until $\overline{Q}$ is not empty, we pick randomly a $v \in \overline{Q}$, retrieve its $MDST(v)$ that we already computed, add its vertices to the connector $S$, and update $\overline{Q}$ by removing the terminal nodes which are reachable from $v$.

The details of the second phase of our method are reported in Algorithm~\ref{algorithm:MIS},
lines \ref{line:transitive_closure}-\ref{line:end_phase2}.

\subsection{Phase 3: Minimum Temporal-Inefficiency Subgraph}
During the second step we have created a connector, i.e., a set of vertices $S \supseteq Q$. Following \cite{RuchanskyBGGK17}, in the third and last step our method greedily removes from $S \setminus Q$ the vertex $v^*$ which provides the greatest improvement in the objective function.

The process is iteratively continued until  $S \setminus Q$ is not empty.
In the end all the intermediate subgraphs produced during the process are compared and the one which provides the smallest temporal inefficiency is returned as  solution.
The third phase of our method is described in Algorithm~\ref{algorithm:MIS},
lines \ref{line:greedy_begin}-\ref{line:greedy_end}.

\begin{algorithm}[h!]
	\caption{Minimum Temporal-Inefficiency Subgraph}\label{algorithm:MIS}
	\Input{Temporal graph $G_W = (V, E_W)$,  $\alpha \in [0,1]$,   $Q \subseteq V$}
	\Output{$G_W[S]$ such that $Q\subseteq S \subseteq V$}
		
	$Q'$ \define $\{(u,-1) | u \in Q\}$\\\label{line:begin_transformation}
   $V_W$ \define $\{(v,t) | t \in W, v \in V\}$ \\
   $V'$  \define $V_W \cup Q' \cup Q$\\

   $E'$ \define $\{((v,t_i),(v,t_{i+1})) | t, t+1 \in W, v \in V \}$\\
   \textbf{for} {$e \in E'$} \textbf{do} { $\ell(e)$ \define $1 - \alpha$}\\
 \textbf{for} {$e \in E_W$} \textbf{do} { $\ell(e)$ \define $\alpha$}\\
    $E'$ \define  $E' \cup E_W$

	$G'$ \define $(V',E',\ell)$\label{line:end_transformation} \\
	$\mathbb{G}'$ \define Transitive closure of $G'$\label{line:transitive_closure}
	
	\For{$r \in Q$\label{line:start_stTree}}
	{
		$MDST(r)$ \define MinimumDST($\mathbb{G}'$, $r$, $Q'$)
	}
$r^*$ \define $\argmin_{r \in Q} \ell(MDST(v))$\\
$S$ \define   $\{ u \in V| (u,t_i) \in MDST(r^*) \}$\\
\If{$\ell(MDST(r^*)) = \infty$}{
$\overline{Q}$ \define $\{(u,-1) \in Q' | d_{MDST(r^*)}(r^*,(u,-1)) = \infty \}$\\
\While{$\overline{Q} \neq \emptyset$}{
pick $v$ from $\overline{Q}$\\
$S$ \define   $S \cup \{ u \in V| (d_{MDST(v)}(v,(u,-1)) \neq \infty \}$\\
$\overline{Q}$ \define $\overline{Q} \setminus \{(u,-1) \in \overline{Q} | d_{MDST(v)}(v,(u,-1)) \neq \infty \} $\label{line:end_phase2}
}
}
$mincost$ \define $\mathcal{I}(G_{W}[S])$\\ \label{line:greedy_begin}
$S^*$ \define $S$\\
\While{$S \setminus Q \neq \emptyset$}{
$v^*$  \define $\argmax_{v \in S \setminus Q} \mathcal{I}(G_{W}[S]) -  \mathcal{I}(G_{W}[S \setminus \{v\} ])$\\
$S$ \define $S \setminus \{v^*\}$\\
\If{$\mathcal{I}(G_{W}[S \setminus \{v\}]) < mincost$}
{$mincost$ \define $\mathcal{I}(G_{W}[S \setminus \{v\}])$\\
$S^*$ \define $S \setminus \{v\}$
}
}
\Return{$G_{W}[S^*]$}\label{line:greedy_end}
\end{algorithm}


\subsection{Streaming Distributed Implementation} \label{sec:distributed}
In the previous sections we have presented in details our method, summarised in Algorithm~\ref{algorithm:MIS}, for the case of a temporal graph $G_W$ defined over a fixed temporal window $W$.
In the more general case, we consider an infinite stream of input graphs. At each new timestamp $t$ the incoming graph $G_t$ becomes the latest snapshot of the \emph{sliding window }$W$, which maintains its size stable by leaving out the most obsolete snapshot. Each time the window $W$ gets updated, we apply Algorithm~\ref{algorithm:MIS} to the new $G_W$ to produce a new minimum temporal-inefficiency subgraph. After this, the query set $Q$ is potentially updated by using the update rules presented in Section 3. We deployed our framework in a streaming distributed implementation on Apache Spark.

The main computational bottleneck of Algorithm~\ref{algorithm:MIS} is at line \ref{line:transitive_closure}:
in fact, the algorithm from~\cite{HuangFL15} requires in input the transitive closure of the graph. Producing the transitive closure requires some computationally expensive algorithm, such as Floyd-Warshall, which has $\mathcal{O}(V^3)$ complexity. In order to avoid this step, we adopt a lazy evaluation approach computing on the fly the transitive closure between two vertices in the graph, when and only if this is required. Instead of using Floyd-Warshall algorithm, we use Dijkstra's algorithm, and we calculate the shortest paths from a given source vertex to the rest of the reachable vertices of the graph. In this way we avoid computing the shortest paths from the vertices that are unreachable temporally from the query vertices. The computation of all shortest paths from each query vertex is done in parallel. Depending on the depth $i$ that is used as a parameter for the computation of the minimum DST as described by~\cite{HuangFL15}, we continue $i$ times the process by choosing as source vertices the $i$-th step neighbours of the query vertices in $Q$. Finally, we distribute the computation of the minimum DST that uses as root each query vertices in $Q$ (lines 10-11 of  Algorithm~\ref{algorithm:MIS}).
%

\section{Experiments}
\label{sec:experiments}

In our experiments we use three real-world dynamic networks summarized in Table~\ref{Table:Dynamic_Graphs_adaptive}.
Two datasets are face-to-face interaction networks gathered by the SocioPatterns\footnote{\url{http://www.sociopatterns.org}} project  using wearable proximity sensors in schools, with a temporal resolution of $20$ seconds.
\textsf{PrimarySchool} contains the contact events between $242$ volunteers ($232$ children and $10$ teachers) in a primary school in Lyon, France, during two days. The data expand across 19 hourly timestamps. \textsf{HighSchool} describes the close-range proximity interactions between students and teachers ($327$ individuals overall) of nine classes during five days in a high school in Marseilles, France. The data expand across 41 hourly timestamps.

The third dataset is the co-authorship network of 16 conferences and journals from the databases, data mining, and information retrieval areas (VLDB, SIGMOD, ICDE, EDBT, KDD, ICDM, SIGIR, CIKM, WWW, WSDM, ECIR, ECML, TKDE, TODS, IEEE BigData and Data Mining and Knowledge Discovery) collected from DBLP\footnote{\url{http://dblp.uni-trier.de/}}. Each vertex is an author and each edge represents co-authorship. It contains 18 yearly timestamps from 2000 to 2017.

\spara{Experimental environment.}
We run our experiments in on a 3.1 GHz Intel Core i7 machine with 16 GB of memory. We use local mode Spark execution with four worker nodes. In each worker we allocate one executor.
For the computation of the DST we set the recursion depth to 1.

\begin{table}[h!]
	\caption{Characteristics of the datasets used.\label{Table:Dynamic_Graphs_adaptive}}\vspace{-2mm}
	
	\centering
	\begin{tabular}{l r r r r}
		Network	& $|V|$	& $|E|$	& $|T|$ & time  granularity\\
	\toprule
		\textsf{PrimarySchool}  				&242					&	7\,420		& 19		&1 hour	\\   
		\midrule
		\textsf{HighSchool}  				&327					&	40\,896		& 41		&1 hour	\\   
		\midrule
		\textsf{DBLP} 						& 46\,160			&377\,852			& 18		&1 year\\ 
		\bottomrule
	\end{tabular}
\end{table}

\subsection{Effect of Parameter $\alpha$}
One of the key notions of our work is that of shortest fastest path (Definition \ref{SFP}). Instead of choosing  between  measuring distances in space or time, we propose to use a linear combination of the temporal and spatial distance, governed by a user-defined parameter $\alpha  \in [0,1]$.
Depending on the application at hand, one can tune the parameter $\alpha$ to give more importance to the temporal dimension ($\alpha < 0.5$) or the spatial one ($\alpha >0.5$), with the extreme cases  $\alpha=0$ and $\alpha =1$ corresponding to fastest-path distance and shortest-path distance, respectively. How to set $\alpha$ might depend on the application at hand, the sensibility of the analyst and the intended semantics of the cohesiveness measure. When there is no reason to favor space over time, nor the other way around, one can simply adopt the default value of 0.5.

In the first set of experiments we analyse the effect of $\alpha$. More in details we ask to what extent the communities found are similar, if we use $\alpha=0.1$ and $\alpha =0.9$. In these experiments we pick a set of query vertices $Q$ at random, with $|Q| = 4$, and we measure the Jaccard similarity between the community returned at every timestamp for the two values of $\alpha$. We use $\lambda_{in} = 1$ (i.e., nodes that are added to $Q$ to form the solution $S$ at time $t$, become part of $Q$ at time $t+1$), while we never drop vertices from $Q$. We repeat each experiment for 100 different $Q$ and report the average.

\begin{figure}[h!]
\vspace{-1mm}
	\centering
\begin{tabular}{cc}
\hspace{-2mm}\includegraphics[width =0.48\linewidth]{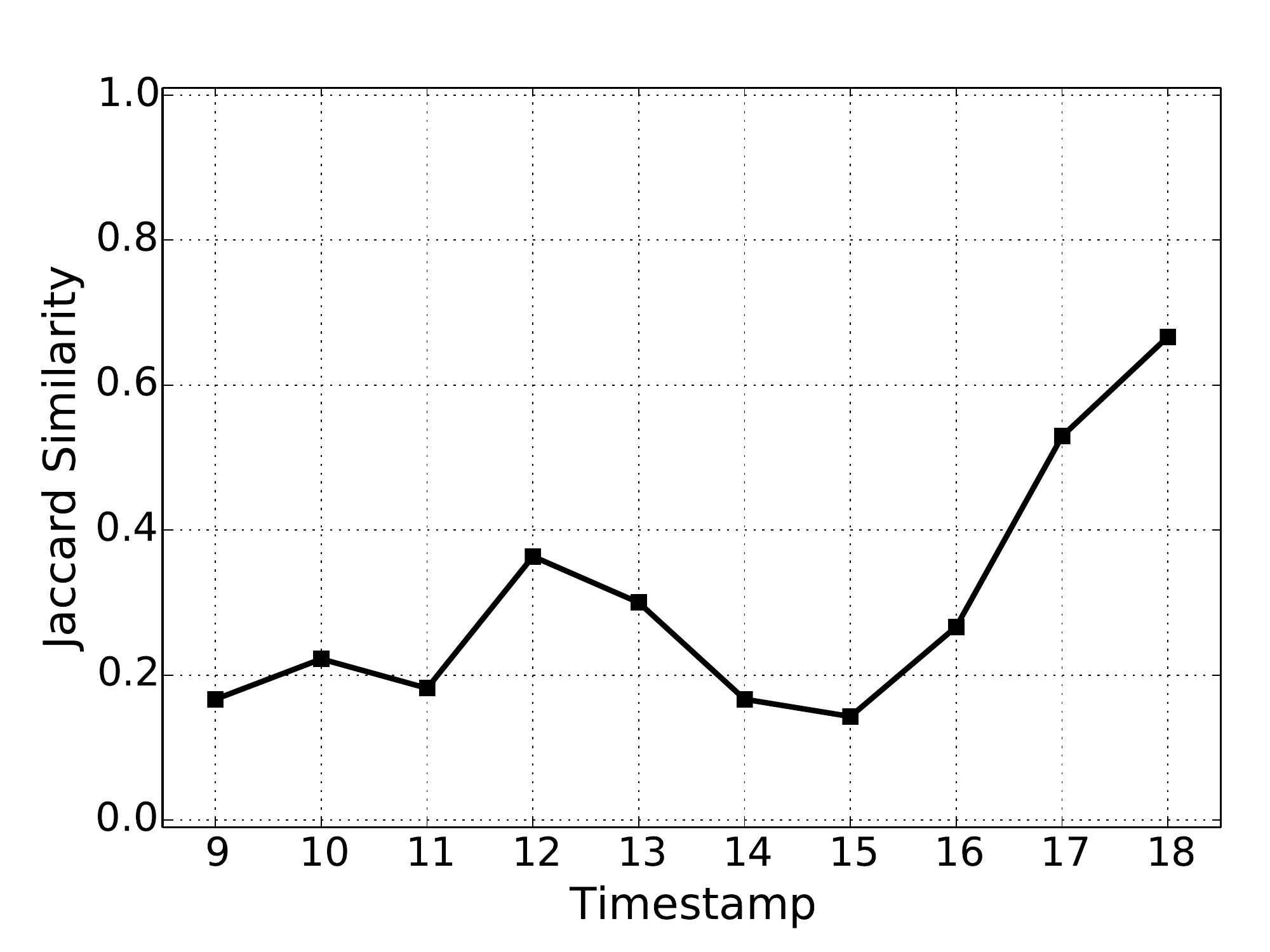} &
\hspace{-2mm}\includegraphics[width =0.48\linewidth]{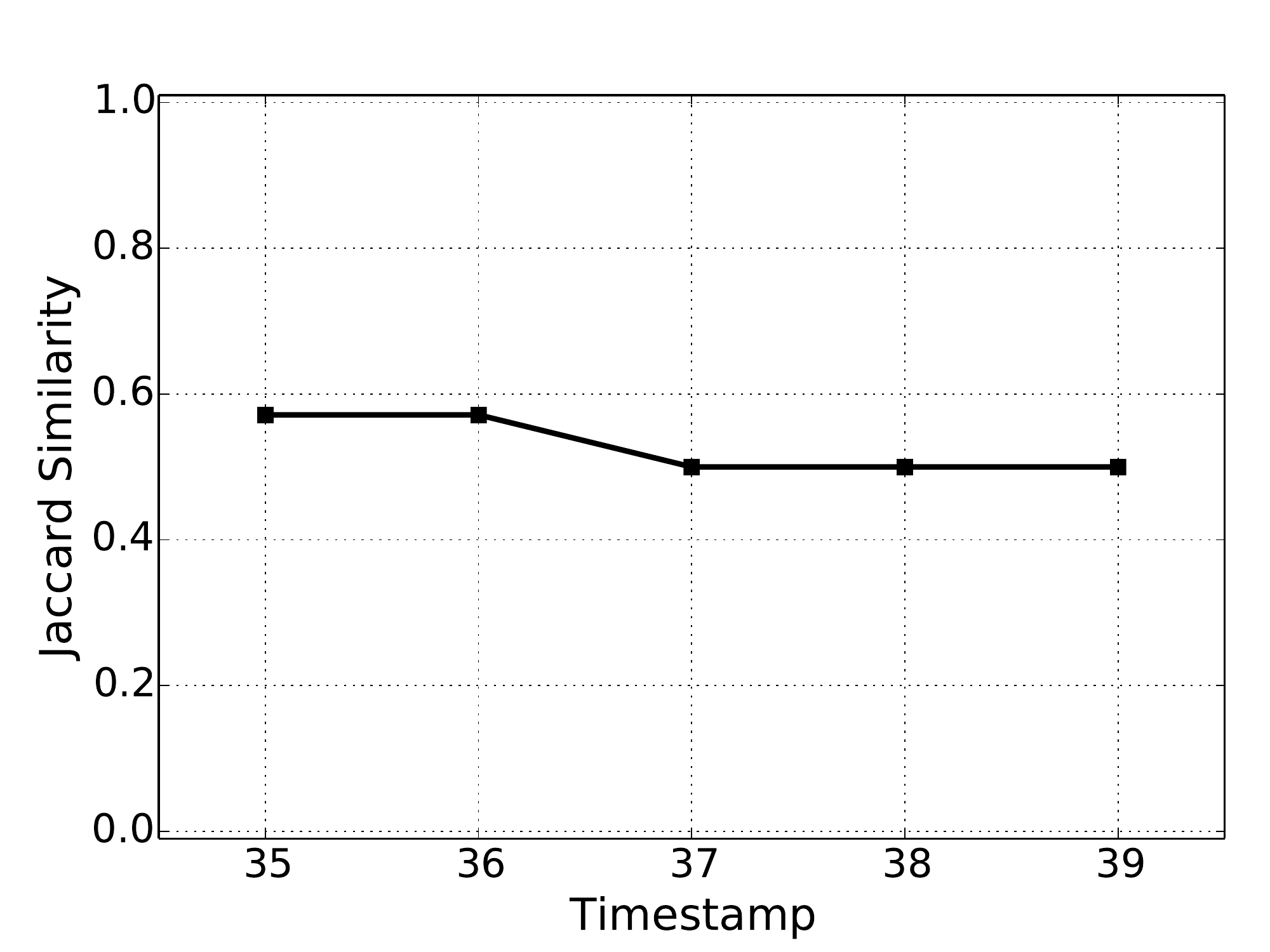}\\
(a) & (b)
\end{tabular}
\caption{Jaccard similarity for the communities detected for $\alpha=0.1$ and $\alpha=0.9$. (a): \textsf{PrimarySchool} with window length $|W|=3$; (b): \textsf{HighSchool} dataset with $|W| = 6$. \label{fig:JaccardPrimarySchool}}
\vspace{-1mm}
\end{figure}

Figure~\ref{fig:JaccardPrimarySchool} (a) shows the Jaccard similarity for \textsf{PrimarySchool} when we set the window size to $|W|=3$, presenting results for $10$ consecutive timestamps. Since the window has size $3$ at each timestamp we see the results that correspond in $3$ hour interval. Therefore, at timestamp $14$, we see the results of the temporal connector that corresponds to the window $W=[12,14]$. The Jaccard similarity varies between 0.18 to 0.7 depending on the timestamp. Confirming that different values of $\alpha$ produce different behaviors.
However,  as the network evolves, the similarity between the solutions returned with $\alpha=0.1$ and $\alpha =0.9$ keeps growing. This is probably due to the adaptivity of the query set: as important vertices are added to the query set $Q$, the community search becomes more stable, converging on the truly important vertices that relate to the initial set of query vertices.

Figure~\ref{fig:JaccardPrimarySchool} (b) shows the Jaccard similarity for \textsf{HighSchool} when we set the window size to $|W|=6$, reporting results for five consecutive timestamps. Since each snapshot of the dataset corresponds to one hour data, the results reported correspond to 6 hour window interval. Therefore, at timestamp 35, we see the results that correspond to $W=[30,35]$, i.e., starting from the 30th hour until the end of the 35th hour. Here the results are more stable with Jaccard similarity at about 0.5.

\begin{figure}[h!]
	\centering
\includegraphics[width =0.47\textwidth ]{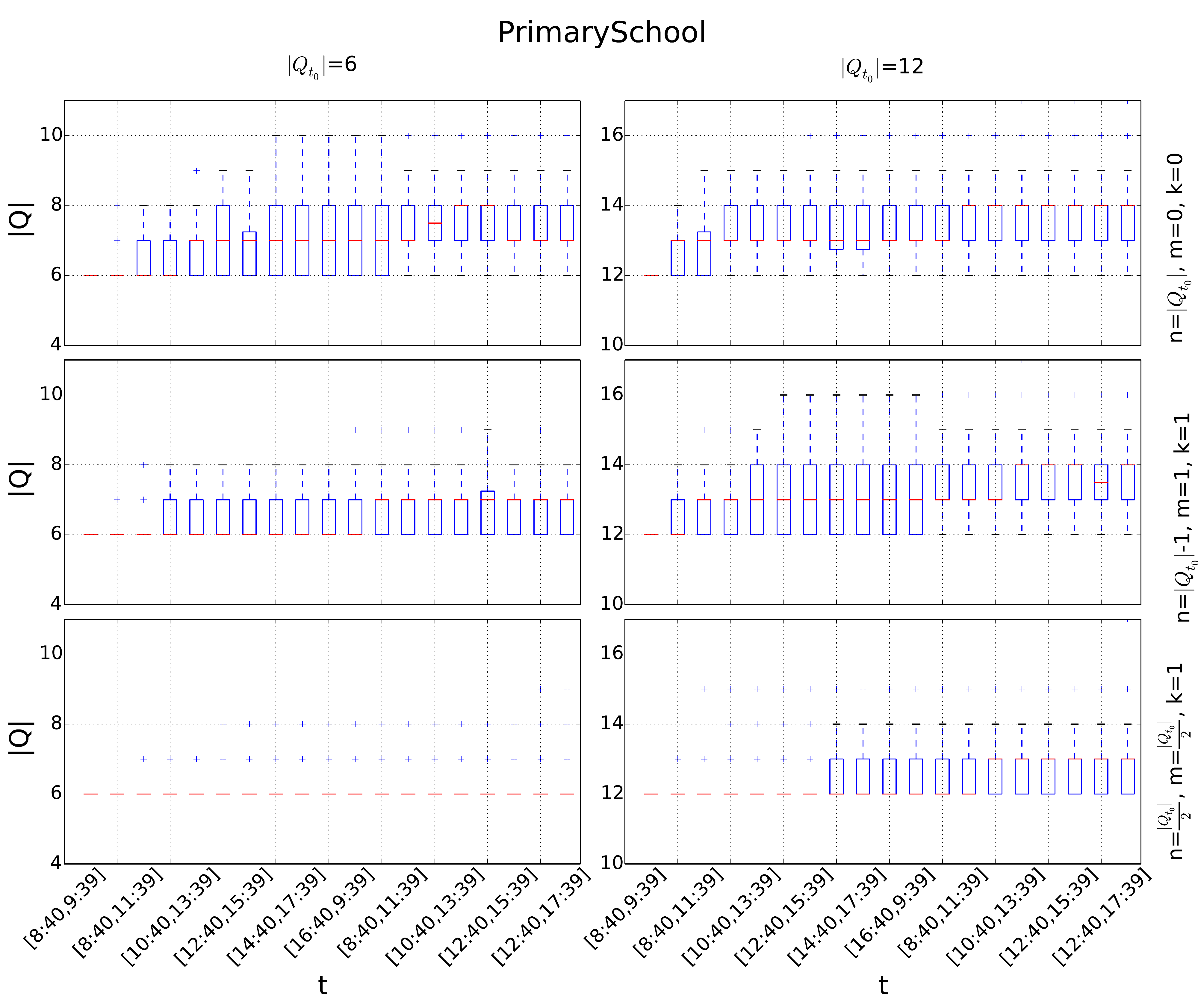}\\
\includegraphics[width =0.47\textwidth ]{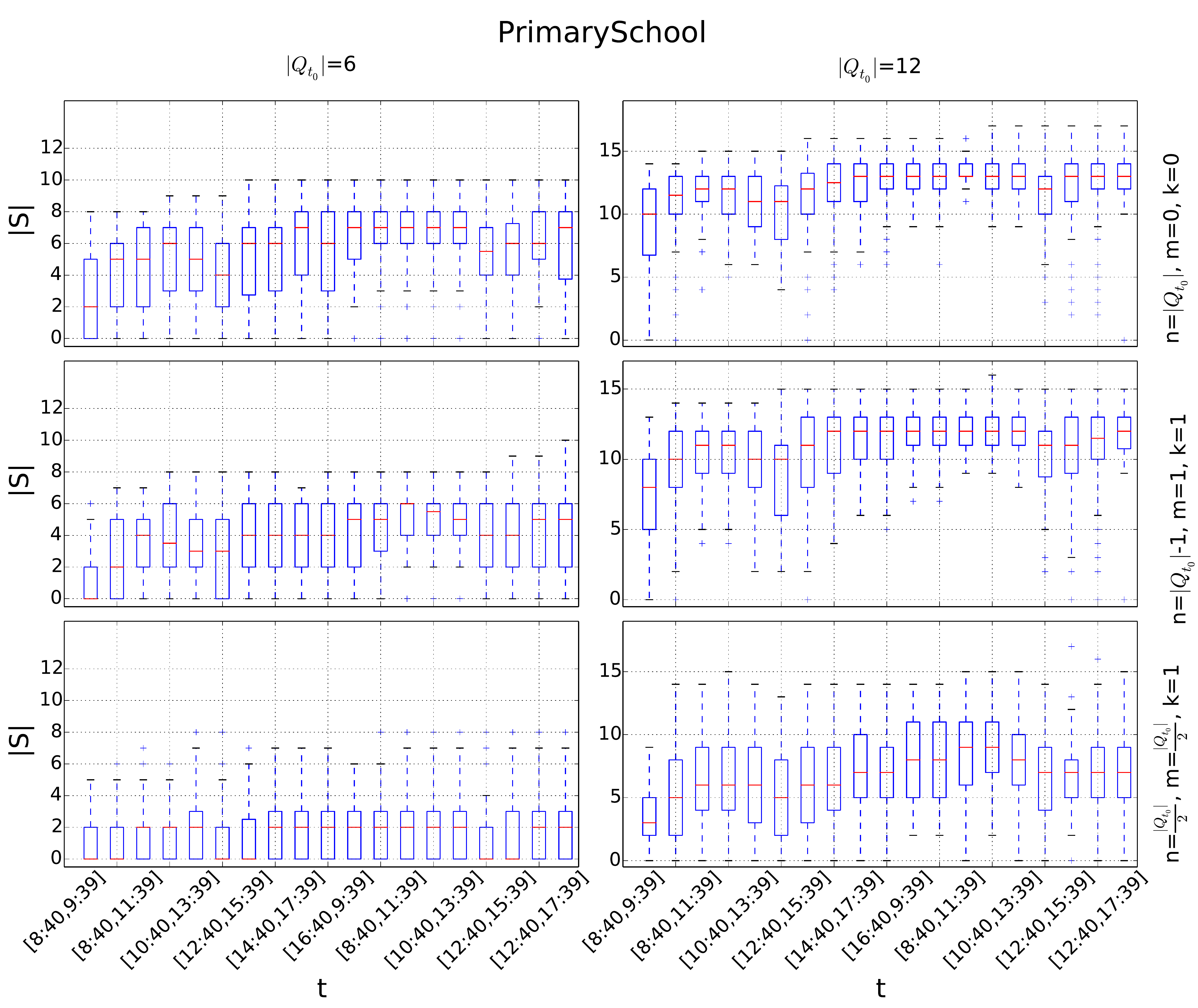}\\
\includegraphics[width =0.47\textwidth ]{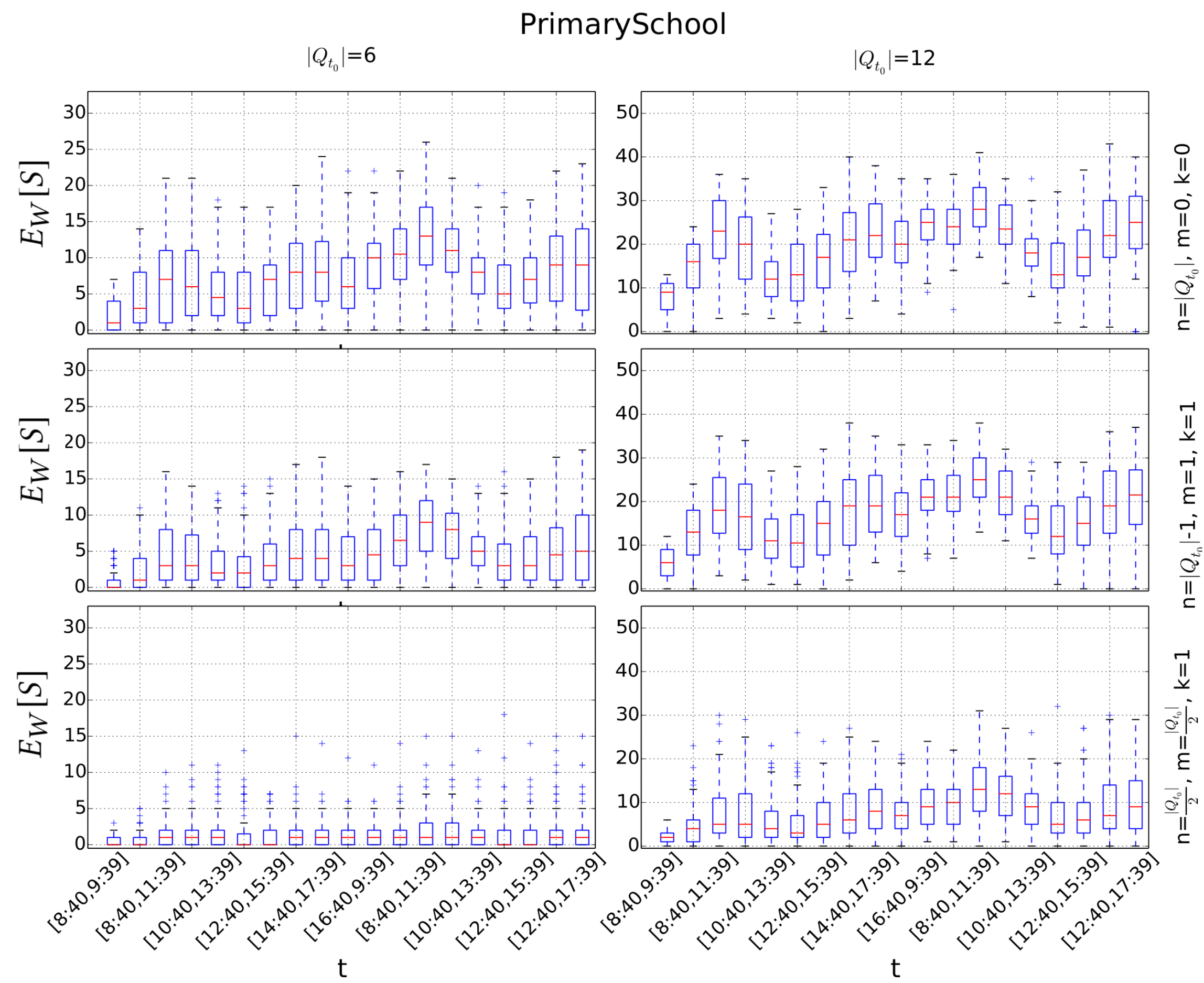}\\

\caption{$|Q|$, $|S|$, and $E_W[S]$ as a function of $t$ for 100 random query sets of size $|Q_{t_0}|= 6$ and 12 for \textsf{PrimarySchool} dataset. The initial query sets $Q_{t_0}$ are selected at random for $(n,m,k)=[(|Q_{t_0}|,0,0), (|Q_{t_0}|-1,1,1), (\frac{|Q_{t_0}|}{2},\frac{|Q_{t_0}|}{2},1)]$.}\label{fig:plot1}
\end{figure}


%

\subsection{Solutions Characterization}
We next characterize the solutions produced along time.
For each experiment we present aggregated results for 100 randomly selected query sets.
Following \cite{RuchanskyBGGK17} se select the initial query set, which in the rest of this section will be denoted $Q_{t_0} \subseteq V$, by using three different parameters:
\squishlist
\item  the size of the initial query set $|Q_{t_0}|$,
\item the number of vertices $n < |Q_{t_0}|$ that we select from a randomly chosen ``main" community,
\item the number of communities $k$ from which we select the $m = |Q_{t_0}| - n$ remaining vertices.
\squishend

This way we can control the number of ``outliers" that we know are contained in $Q_{t_0}$.
More specifically,  we select at random a community from which we extract $n$ vertices. The other $m= |Q|-n$ vertices are extracted from $k$ other communities. Notice here that when $m=k$, it is implied that we have $n$ vertices from the same community and $m$ outliers. For this set of experiments we use \textsf{PrimarySchool} and \textsf{HighSchool} datasets, for which we already know the underlying communities, i.e., students belonging to the same class form a community. We produce 100 query sets for each one of the following three combinations of the parameters $n$, $m$, $k$:
\squishlist
\item no outliers, i.e., $(|Q_{t_0}|,0,0)$;
\item one outlier, i.e., $(|Q_{t_0}|-1,1,1)$;
\item query set split in two communities, i.e., $(|Q_{t_0}|/2,|Q_{t_0}|/2,1)$.
\squishend
We repeat out experiments for values of $|Q_{t_0}|=6$ and $12$. As before, we use $\lambda_{in} = 1$ (i.e., nodes that are added to $Q$ to form the solution $S$ at time $t$, become part of $Q$ at time $t+1$), while we never drop vertices from $Q$. Figure~\ref{fig:plot1} reports the results for \textsf{PrimarySchool}:  we focus on the size of the query set, $|Q|$, at the beginning of each timestamp, the size of the solution in terms of number of vertices $|S|$ and number of induced temporal edges $|E_W[S]|$.

We observe that all the different characteristics of the temporal connector ($|Q|, |S|$ and $|E_W[S]|$) exhibit the same behavior. In the first setting (first row plots), where all the vertices are concentrated in the same community, the size of the query set increases as the time evolves. However, in the second setting (second row plots), this behavior is smoothed as we notice a smaller increase of the size of $Q$ that starts in a later timestamp. Finally, in the third setting, where the vertices are equally partitioned in the two communities, there is almost no increase in the query set for $|Q_{t_0}|=6$ and  a small increase for $|Q_{t_0}|=12$ starting at timestamp $7$. These results verify our hypothesis that our approach connects vertices that exist in the same community, and adds parsimoniously vertices in the temporal connector. We additionally notice, that thanks to the adaptivity of the query set the increase and the decrease of $|Q|, |S|$ and $|E_W[S]|$ are smooth in time.

\begin{figure*}[t!]
	\centering
	\includegraphics[width =\textwidth ]{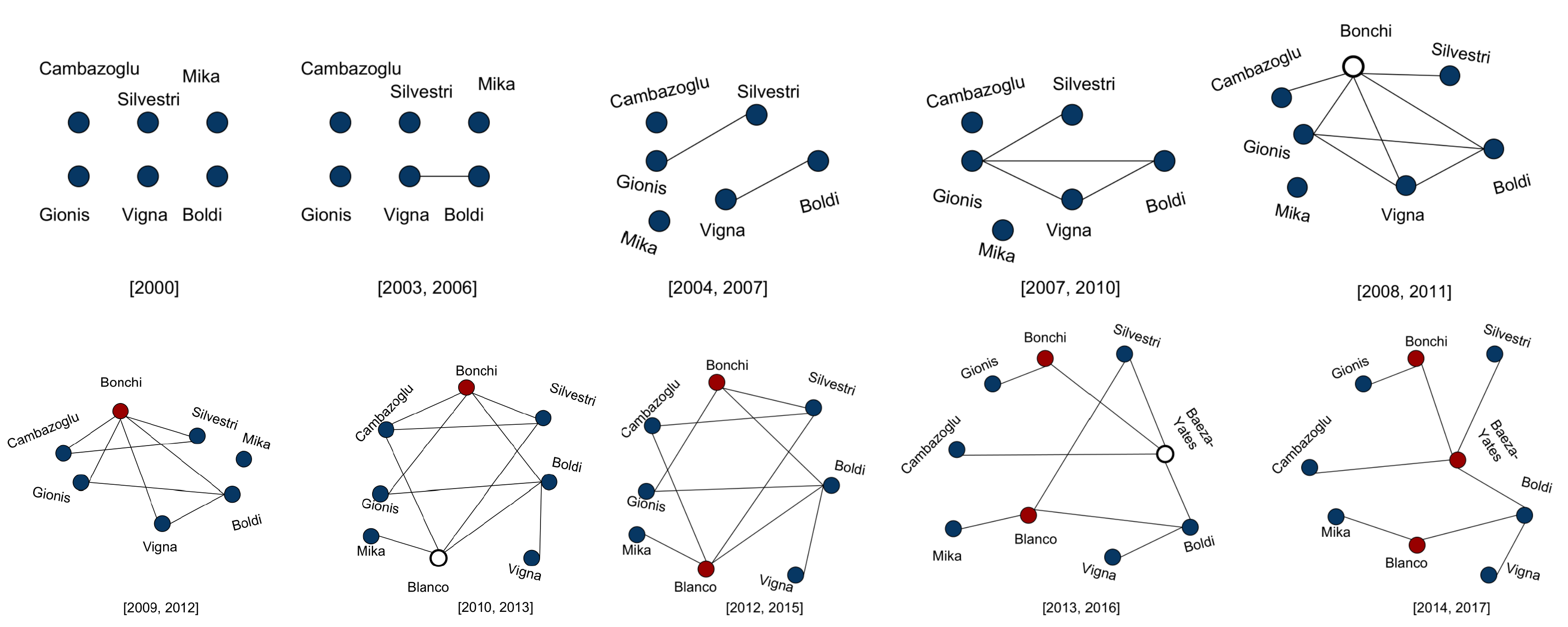}
	\caption{Case study for \textsf{DBLP} dataset. Initial query set $Q_{t_0}$ marked with blue color.
The vertices which form the temporal community with $Q$ are marked in white, while the vertices added to $Q$ are marked in red. In this example $|W|=4$, $\alpha = 0.1$, and  $\lambda_{in} = 1$. The edges are the temporal edges that exist in at least one timestamp within the given time window. The figure reports only some of the }\label{fig:example2}
\vspace{-2mm}
\end{figure*}


We see similar trends for \textsf{HighSchool} data (not reported for sake of space).

%

%


\subsection{Case Study}
\label{sec:casestudies}

Finally we present some anecdotal evidence to give an idea of the practical applicability of our framework in the discovery of the community around a given set of vertices and monitor its evolution along time. Figure \ref{fig:example2} reports an example in the \textsf{DBLP} co-authorship temporal network. The initial query set $Q_{t_0}$ contains six researchers, spanning different research communities, but which were all linked to Yahoo Labs in Barcelona: either employees or frequent academic visitors.

We can observe the following temporal dynamics. Back in time no collaboration was occurring among the six query vertices and the produced solution just leave all of them disconnected. Later on, as the lab in Barcelona is established, collaborations start appearing. However, at the beginning  the solution subgraph is still sparse and formed by only the query vertices. As time goes by, more and more collaborators start being involved and they are added to the solution, as they help making the subgraph more cohesive (reducing the distances among the query vertices). The vertices (marked in white) which are used to produce a minimum temporal inefficiency subgraph, are then added to the query set for the future timestamps (and marked in red).

\section{Conclusions}
\label{sec:conclusions}
In this paper we tackled the problem of community search in temporal dynamic networks and proposed the problem of extracting the minimum temporal-inefficiency subgraph. We developed a method based on a careful transformation of the temporal network to a static directed and weighted graph, and some recent approximation algorithm for the extraction of Directed Steiner Tree.  We provided a streaming distributed implementation of our framework in Apache Spark and experimented on several real-world networks.

While the community search problem is well studied, both the ``relaxed" variant, which allows some query vertices to remain disconnected, and the variant on temporal networks are largely and surprisingly unexplored. These are important problems which, in the coming years, will receive more foundational treatment and will find more applications. This work represents just a first step in this direction.

\bibliographystyle{abbrv}
\bibliography{referencesTKDD,references,refs_short,biblio_short}

\end{document}